\documentstyle[aps,prb,multicol,epsf]{revtex}

\begin{document}
\draft

\title{Analysis of negative magnetoresistance. \\
Statistics of closed paths. I. Theory}
\author{G. M. Minkov, A. V. Germanenko, \cite{cont}
V. A. Larionova  and S. A. Negashev}
\address{Institute of Physics and Applied Mathematics, Ural
State University \\
620083 Ekaterinburg, Russia}
\author{I. V. Gornyi}
\address{A. F. Ioffe Physical-Technical Institute, 194021 St. Petersburg,
Russia}
\date{\today}
\maketitle \widetext
\begin{abstract}
Statistics of closed paths in two-dimensional (2D) systems, which just
determines the interference quantum correction to conductivity and anomalous
magnetoconductance, has been studied by computer simulation of a particle
motion over the plane with randomly distributed scatterers. Both ballistic and
diffusion regimes have been considered. The results of simulation have been
analyzed in the framework of diffusion approximation. They are used for
calculation of the magnetic field dependence of magnetoconductance in the
model 2D system. It is shown that the anomalous magnetoconductance can be in
principle described by the well known expression, obtained in the diffusion
approximation, but with the prefactor less than unity and phase breaking
length which differs from true value.
\end{abstract}


\begin{multicols}{2}

\narrowtext
\section{Introduction}
\label{sec:itro} It is well known that the interference of electron waves
scattered along closed trajectories in opposite directions produces a quantum
correction to the conductivity. An external magnetic field applied
perpendicular to the two-dimensional (2D) layer destroys the interference and
suppresses the quantum correction. This results in anomalous negative
magnetoresistance, which is experimentally observed in many 2D systems. This
phenomenon can be described in the framework of quasiclassical approximation
which is justified under the condition $k_F l\gg 1$, where $k_F$ is the Fermi
wave vector, $l$ is the mean free path. In this case the conductivity
correction is usually expressed through the classical quasiprobability for an
electron to return to the area of the order $\lambda_F l$ ($\lambda_F=2\pi
/k_F$) around the start point \cite{gork,chak,dyak,dmit}
\begin{equation} \label{eq1}
\delta\sigma=-\sigma_0 \frac{\lambda_F l}{\pi} W,
\end{equation}
where $\sigma_0=e^2 k_F l/(2\pi\hbar)$, and $W$ stands for the
quasiprobability density of return ({\em quasi-} means that $W$ includes not
only the classical probability density, but the effects of interference
destruction due to an external magnetic field and inelastic scattering
processes). In order to calculate a magnetic field dependence of negative
magnetoresistance the quasiprobability $W$ is represented as a sum of
contributions of closed paths with $N$ collisions, $W_N$. Then, Eq.\
(\ref{eq1}) can be rewritten as
\begin{equation}\label{eq1a}
\delta\sigma= -2\pi l^2 G_0 \sum_{N=3}^{\infty}W_N,
\end{equation}
where $G_0=e^2/(2\pi^2\hbar)$. Here, only paths with $N\geq 3$ are taken into
account, because the paths with $N=1,\,2$ have zero areas and their
contributions are not influenced by the magnetic field.

The expression (\ref{eq1a}) for backscattering quantum correction is true for
an arbitrary magnetic field, any anisotropy of scattering and distribution of
scatterers, and various relationship between phase and momentum relaxation
times, $\tau_\varphi$ and $\tau$, respectively. The sum (\ref{eq1a}) is
usually calculated by means of diagrammatic technique. \cite{schm,hik}
Analytical expressions for negative magnetoresistance have been obtained this
way for random distribution of scatterer in the following cases (i) arbitrary
scattering anisotropy for low magnetic field $B<B_{tr}$,\cite{aniz} where
$B_{tr}=\hbar c/(2e l^2)$; (ii) isotropic scattering  for $B\gg B_{tr}$.
\cite{dmit} In the diffusion approximation, i. e. when the number of
collisions for actual trajectories is much greater than unity, this procedure
gives \cite{schm}
\begin{eqnarray}
\label{eq2} \Delta\sigma(b)&=&\delta\sigma(b)-\delta\sigma(0)\nonumber \\ &=&a
G_0 \left[ \psi\left(\frac{1}{2}+\frac{\gamma}{b}\right)-
\psi\left(\frac{1}{2}+\frac{1}{b}\right)- \ln{\gamma} \right],
\end{eqnarray}
where $\gamma=\tau/\tau_\varphi$, $b=B/((1+\gamma)^2 B_{tr})$, $\psi(x)$ is a
digamma function, and  $a$ is so called prefactor, which is equal to unity
according to the theory. For $x\gg 1$ $\psi(1/2+x)\simeq\ln(x)$, and the
expression (\ref{eq2}) coincides with that obtained in Ref.\ \onlinecite{hik}.
The calculations of $\delta\sigma(b)$ beyond the diffusion limit
\cite{dyak,schm} show that $\delta\sigma(b)$ markedly deviates from this
theory if the number of collisions for actual trajectories is not very large.
The role of nonbackscattering contribution to magnetoconductance has been
studied in Ref.\ \onlinecite{dmit}. This contribution has been found to cause
the reduction of scattering at arbitrary angles and, in contrast to the
coherent backscattering, the conductivity increasing. In the diffusion limit
the nonbackscattering contribution is negligible small, but in the case of a
strong magnetic field $B>B_{tr}$ it should be taken into account.

It is usual to analyze experimental data by means of equation (\ref{eq2}). If
this equation describes the magnetic field dependence of negative
magnetoresistance satisfactorily, it is possible to determine $\tau _{\varphi
}$  and its temperature dependence.

In our two papers presented back-to-back we put forward a new approach to
calculation and analysis of negative magnetoresistance. By representing the
quasiprobability $W$ as a sum of contributions from trajectories with given
areas we express the negative magnetoresistance in terms of area distribution
function of closed paths $W(S)$ and area dependence of their average lengths
$\overline{L}(S)$. It is shown that these are precisely the statistic
characteristics which can been obtained from the analysis of experimental data
(see the following paper). In the present paper the statistics of closed paths
is studied theoretically by using computer simulation. This method allows to
obtain the statistic characteristics of closed paths beyond the diffusion
approximation without any restriction on the scattering anisotropy and
impurity distribution when analytical expressions cannot be derived.

This paper is organized as follows. In the next section we give the necessary
formulas and definitions. In Section \ref{sec:sdet} the details of simulation
procedure are presented. The statistics of closed paths obtained from the
simulation is  given in Section \ref{sec:stat}. The results are compared with
those obtained in the framework of the diffusion theory. In Section
\ref{sec:nmr} the magnetic field dependence of negative magnetoresistance of
the model 2D system are presented and analyzed. Both coherent backscattering
and nonbackscattering contributions to magnetoconductance are considered.

\section{Basic equations}
\label{sec:nmres} Let us introduce the value $w_N(S)$ in such a way that
$w_N(S)d S$ gives the probability density of return after $N$ collisions
following a trajectory, which encloses the area in the range $(S,S+dS)$. In
this case Eq.\ (\ref{eq1a}) for conductivity correction in a magnetic field is
written as follows
\begin{eqnarray}
  \delta\sigma(b)&= &-2\pi l^2 G_0 \times \nonumber \\
    &\times &\sum_{N=3}^{\infty}\int_{-\infty}^{\infty}
  d S\, w_N(S)  \cos \left(\frac{(1+\gamma)^2 b S}{l^2}\right).
  \label{eq502}
\end{eqnarray}
In order to take into account inelastic processes destroying the phase
coherence we include the factor $\exp(-L/l_\varphi)$ in Eq.\ (\ref{eq502}),
where $l_\varphi$ is the phase breaking length connected with $\tau_\varphi$
through the Fermi velocity, $l_\varphi=v_F\tau_\varphi$ and replace the
summation over $N$ by integration over the path length $L$. Then, Eq.\
(\ref{eq502}) takes the form
\begin{eqnarray}\label{eqSbL}
  \delta\sigma(b)&=& -2\pi l^2 G_0 \int_0^{\infty} \frac{d L}{l}\biggl\{
 \exp\left(-\frac{L}{l_\varphi}\right)\nonumber \\
 & &\int_{-\infty}^{\infty} dS\ w(S,L) \cos \left(\frac{(1+\gamma)^2 b S}
 {l^2}\right)\biggr\}.
\end{eqnarray}
Here, $w(S,L)d S$ gives the density probability of return along a trajectory
with the length $L$ and area in the interval $(S,S+dS)$. Let us introduce the
average length $\overline{L}$ of closed paths with a given area in such a way:
\begin{eqnarray}\label{eq7}
  \exp\left(-\frac{\overline{L}(S)}{l_\varphi}\right)&=&
  \frac{1}{W(S)}\times\nonumber \\
  &\times&\int_0^\infty \frac{d L}{l}\
  w(S,L)\exp\left(-\frac{L}{l_\varphi}\right),
\end{eqnarray}
where
\begin{equation}\label{eq7a}
  W(S)=  \int_0^\infty \frac{d L}{l} w(S,L).
\end{equation}
In this case Eq.\ (\ref{eqSbL}) can be rewritten as follows
\begin{eqnarray}\label{eq60}
    \delta\sigma(b)&=&-2\pi l^2 G_0 \int_{-\infty}^\infty dS\ \biggl\{W(S)
    \nonumber\\
    & &\exp\left(-\frac{\overline{L}(S)}{l_\varphi}\right)\cos
  \left(\frac{(1+\gamma)^2 b S}{l^2}\right)\biggr\}.
\end{eqnarray}
Thus, the area distribution function $W(S)$ and function $\overline{L}(S)$
play a decisive part in the magnetic field dependence of $\delta\sigma$. It is
clearly seen that these functions can be extracted from experimental
$\delta\sigma(b)$ curves by using Fourier transformation. To understand what
these statistic characteristics are let us turn now to theoretical study of
statistics of closed paths through the computer simulation of a particle
motion over 2D plane with randomly distributed scatterers.

\section{Simulation details}
\label{sec:sdet} 
\begin{figure}
 \epsfclipon
 \epsfxsize=\linewidth
 \epsfbox{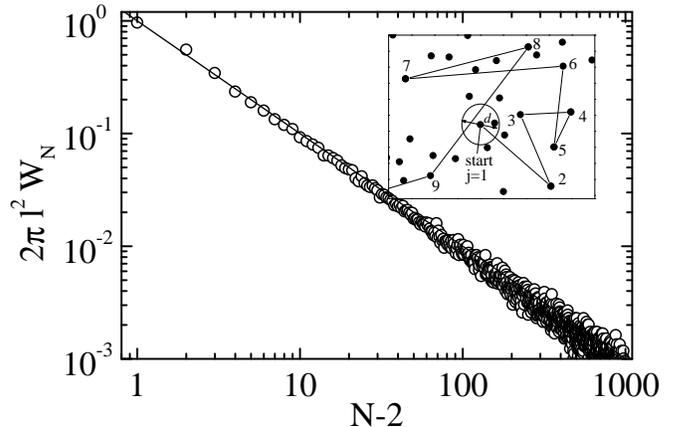}
\caption{ The quantity $2\pi l^2 W_N$ as a function of collision number $N$.
Circles are the result of simulation. The line is theoretical calculation
using Eq. (\protect\ref{eq5}). The inset shows schematically one of the paths
which pass through the start point vicinity after $N=8$ collisions.}
\label{fig1}
\end{figure}

The model 2D system is conceived as a plain with randomly
distributed scattering centers with a given total cross-section. It is
represented as a lattice $M\times M$. The scatterers are placed in a part of
the lattice sites with the use of a random number generator. We assume that a
particle moves with a constant velocity along straight lines which happen to
be terminated by collisions with the scatterers.\cite{otsenka}  The following
algorithm has been realized in computer to obtained the information about the
statistics of closed paths:
\begin{enumerate}
 \item \label{sim1}
A start point is chosen to coincide with some scatterer near the center of the
lattice;
 \item \label{sim2}
We consider a particle suffers the first collision in the start point and
begins to move in a random direction (see inset in Fig.\ \ref{fig1});
 \item \label{sim3}
If the particle passes in the vicinity of some scatterer at a distance less
than $s/2$, where $s$ is the total cross-section of the scatterer, the
collision is considered to take place;
 \item \label{sim4}
Then a scattering angle is randomly generated and the particle begins to move
in the new direction until the next collision occurs. The distances between
two sequential collisions are stored in order to calculate the mean free path;
 \item \label{sim5}
If the trajectory of the particle passes near the start point at the distance
less than $d/2$ (where $d$ is a prescribed value, which is small enough), it
is perceived as being closed. Its length $L$, the number of collisions $N$,
the angle $\theta$ between the first and the last segments of the trajectory,
and the enclosed algebraic area, calculated according to
\begin{eqnarray}\label{eqS}
S=\sum_{j=1}^{N-1}\frac{y_{j+1}+y_j}{2} (x_{j+1}&-&x_j)+ \nonumber \\
+\frac{y_{N}+y_1}{2}(x_{N}&-&x_1),
\end{eqnarray}
where $x_j$, $y_j$ stand for coordinates of $j$-th collision, are kept in
memory. Then, the particle continues moving in the same direction. Notice that
multi-returned trajectories are taken into account. Their length and area are
calculated beginning with the start point;
 \item \label{sim6}
Steps \ref{sim3}--\ref{sim5} are repeated until the particle goes out of the
lattice or a number of collisions exceeds some large prescribed value $N_m$;
 \item \label{sim7}
Then, another particle is launched from the start point in a random direction,
and steps \ref{sim3}--\ref{sim6} are repeated;
 \item \label{sim8}
After a thousand of starts, a new start point is chosen, and steps
\ref{sim2}--\ref{sim7} are repeated, for tens of times;
 \item \label{sim9}
Then a new ensemble of scatterers is randomly generated and the whole
procedure is done over and over again.
\end{enumerate}

The procedure described allows to find the probability, ${\cal T}$, for the
trajectory to pass at a distance less than $d/2$ near the start point, instead
of the probability density of return, $W$, which stands in Eq.\ (\ref{eq1}).
But it can be shown (see Appendix) that  for $d\ll l$ these values are
connected through the simple relationship $W=(dl)^{-1}{\cal T}$.

All the results presented in the paper have been obtained using the following
parameters: the lattice dimension is $6500\times6500$, the total number of
scatterers is about $8\times 10^4$, $N_m=10^3$, $s=7$, and $d=1$ (hereafter
all the lengths and areas are given in units of lattice parameter and lattice
parameter squared, respectively). The total number of starts, $I_{s}$, is
about $10^6-10^7$. The mean free path computed for such a system is $77.8$. It
is close to theoretical estimation of mean free path $l=(N_s s)^{-1}$, where
$N_s$ is a density of scatterers. The calculations were carried out also with
other parameters $M$ and $s$ differed $2-3$ times. It has been shown that this
leads only to a change in the value of $l$, but does not influence the
dependence of magnetoconductance on the reduced magnetic field $b=B/B_{tr}$.

In order to illustrate the 2D system which can correspond to our model in
reality, let us set the lattice constant equal to $0.5$ nm. In this case our
model provides an example of 2D system with the concentration of scatterers
$7.5\times 10^{11}$ cm$^{-2}$, mean free path $l=38.9$ nm, and $B_{tr}\simeq
0.2$ T.

\section{Statistics of closed paths}\label{sec:stat}

\subsection{Simulation results}\label{subsec:stat1}

It has been shown in Ref.\ {\onlinecite{dyak}} that if we decompose the
probability density of return to the origin $W$ as a sum of contributions of
paths with $N$ collisions, $W_N$, for each partial contribution a simple
expression is valid:
\begin{equation}\label{eq5}
  2\pi l^2 W_N=(N-2)^{-1},\;  N\ge 3.
\end{equation}
We emphasize that Ex.\ (\ref{eq5}) is exact. It has been obtained for random
distribution of scatterers of zeroth radius. Corresponding simulation results
for our model system are presented in Fig.\ \ref{fig1}. As is clearly seen the
power law works well in the whole range of $N$.  The value of $W_N$ is really
proportional to $(N-2)^{\alpha}$ with $\alpha=(-1.05\pm 0.02)$ which is close
to the theoretical value $\alpha=-1$. The slight difference results from a
specific feature of our system. The matter is that the scatterers are placed
discretely rather than continuously as in the theoretical approach. They
cannot lie closer than one unit cell of the lattice. This leads to the fact
that unlike the theory a distance between two sequential collisions in our
model is always greater than some critical value of the order of lattice
constant. From this point of view, our model system most closely corresponds
to real 2D system, where the ionized impurity distribution is discrete and
correlated to some extend due to Coulomb repulsion at growth temperature. It
is easy to show in the framework of theoretical approach Ref.\
{\onlinecite{dyak}} that the cutoff of short free paths results in more steep
$W_N(N)$ relationship as compared with Eq.\ (\ref{eq5}).

\begin{figure}
 \epsfclipon
 \epsfxsize=\linewidth
 \epsfbox{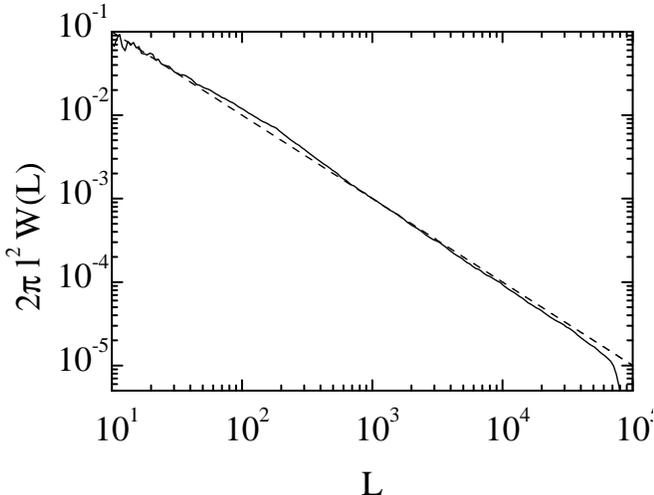}
\caption{Length distribution function of closed paths $W(L)$. The solid curve
is the result of simulation, the dotted curve shows $L^{-1}$ function.}
\label{fig2}
\end{figure}

The length distribution function for closed paths, $W(L)$, is presented in
Fig.\ \ref{fig2}. The function $W(L)$ is defined in such a way that the value
$W (L)d L$ is the probability density of return to the origin following a
trajectory with the length belonging to the interval from $L$ up to $L+dL$. As
is seen $W(L)$ is inversely related to trajectory length. A drastic decrease
of $W(L)$ for $L\gtrsim 7.5\times 10^4$ results from the restriction on
maximal number of collisions per one trajectory in our algorithm. As a
consequence of this restriction the probability of a closed path to have the
length exceeding the value $N_m l\simeq 8\times 10^4$ is negligible small.

\begin{figure}
 \epsfclipon
 \epsfxsize=\linewidth
 \epsfbox{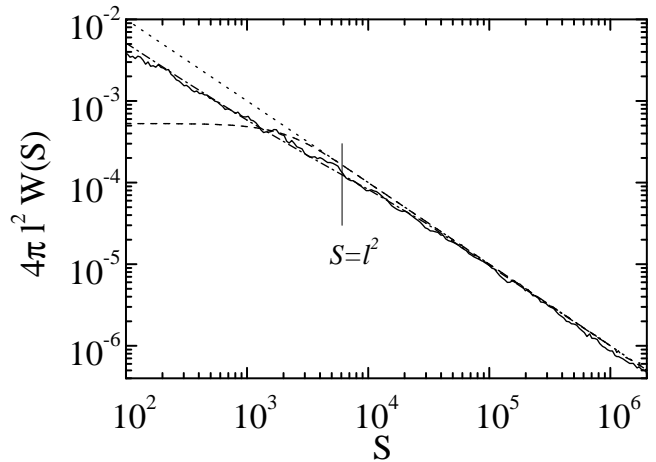}
\caption{Area distribution function of closed paths $W(S)$. The solid curve is
the result of simulation; the dashed and dot-dashed curves are the diffusion
and improved diffusion approximations, respectively; the dotted curve shows
$S^{-1}$ function.}
 \label{fig3}
\end{figure}

Figure \ref{fig3} shows the area distribution function obtained from our
numerical simulation. Since the results of calculations are identical for
positive and negative algebraic areas, hereafter we present theoretical curves
in the positive area range only. It is reasonable that $W(S)$ is a diminishing
function. As is seen it is difficult to describe $W(S)$ curve by a power
function in the whole range of areas. However, for large $S$ the power
function $S^{-1}$ is a good asymptotic of $W(S)$.

\begin{figure}
 \epsfclipon
 \epsfxsize=\linewidth
 \epsfbox{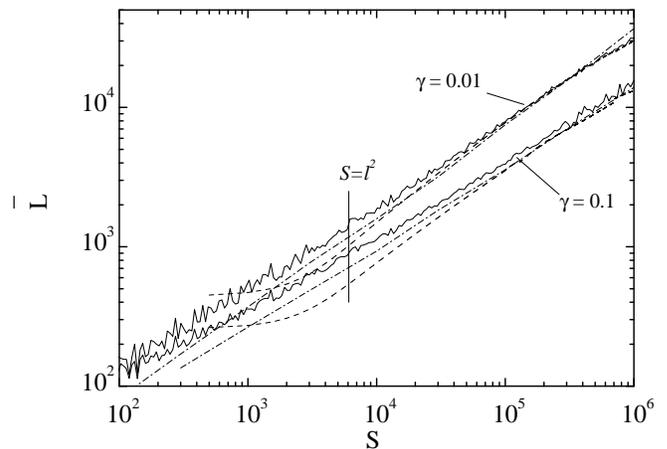}
\caption{The area dependence of $\overline{L}$ for different $\gamma$ values.
Solid curves are the simulation results, dashed and dot-dashed curves are the
results of calculation within the diffusion and improved diffusion
approximations, respectively (see text).}
  \label{fig4}
\end{figure}

The results of our calculation of $\overline{L}(S)$ for different values of
$l_\varphi$ are presented in Fig.\ \ref{fig4}. As is seen the smaller the
value of $\gamma$, i. e. the greater $l_\varphi$, the greater is the average
length of closed paths with a given area. For large area values $S\gtrsim
l^2$, the area dependence of $\overline{L}$ can be described by the power
function $S^\beta$ with $\beta$ slightly depending on the value of $\gamma$.
As $\gamma$ changes from $0.1$ to $0.01$, the value of $\beta$ varies from
$0.55$ to $0.62$.

Although the particle motion in our model 2D system is a special case of the
random walker problem, which is studied well enough, to our knowledge there is
no analytical solution of this problem in a wide range of path lengths and
areas, involving non-diffusion motion. There is no theory which could describe
analytically the statistics of closed paths with small number of collisions.
Below we analyze the simulation results in the framework of diffusion theory.

\subsection{Comparison with diffusion theory}\label{subsec:stat2}

\begin{figure}
 \epsfclipon
 \epsfxsize=\linewidth
 \epsfbox{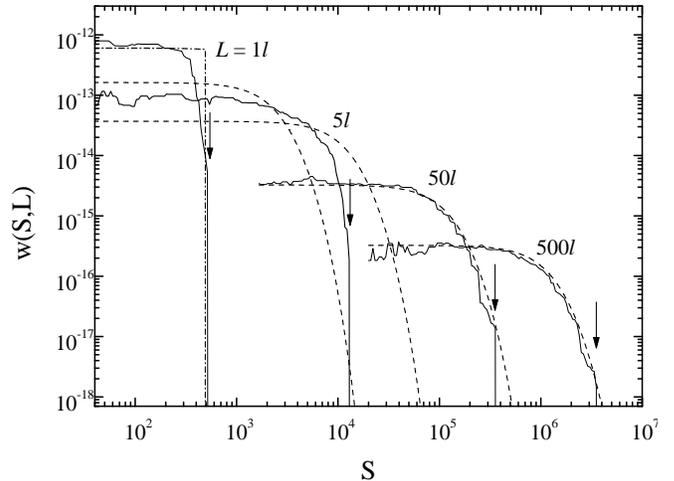}
 \caption{The quantity $w$ as a function of $S$ for
different $L$ values. Solid lines represent the simulation data. Dashed and
dot-dashed lines show the results of calculation using Eqs.\ (\ref{wsl}) and
(\ref{wslImpr}), respectively. Arrows indicate the critical area value as it
is obtained from the simulation procedure.}
  \label{fig41}
\end{figure}

Let us now calculate within the diffusion approximation the quantities
$w(S,L),\ W(S)$, and $\overline{L}(S)$ introduced in Section \ref{sec:nmres}
to describe statistical properties of a random walk. A particle trajectory is
regarded as being diffusive when its length is greater than the mean free
path, associated with the transport scattering time. According to the
diffusion theory the probability density  of finding a particle in the
vicinity of the point ${\bf r}$ at the moment $t$, $P({\bf r},t)$, does not
depend on the initial and final velocity directions and obeys the usual
diffusion equation
\begin{equation}
 \frac{\partial P}{\partial t}-D \Delta P =\delta(t)\delta({\bf r}),
 \label{du}
\end{equation}
where $D$ is the diffusion coefficient. A solution of the diffusion equation
can be written in the form of the Wiener path integral as (see, e.g. Ref.\
\onlinecite{chak})
\begin{equation}
P({\bf r},t)=\int^{{\bf r}(t)={\bf r}}_{{\bf r}(0)=0} {\cal D}{\bf r}(\tau)
\exp\left (-\int^t_0d\tau\frac{{\bf{\dot r}}^2(\tau)}{4D}\right).
\label{wiener}
\end{equation}
With help of this formalism we calculate the probability density ${\cal P}$
for a diffusive walk with a constant velocity $v_F$ and the length $L=v_F t$
to enclose the algebraic area $S$
\begin{displaymath}
{\cal P} (S,L)= \left \langle\delta\left(S-\frac{1}{2}\int^{t=L/v_F}_0d\tau
r^2(\tau){\dot \theta}(\tau)\right)\right \rangle_{P(0,t)}.
\end{displaymath}
Here the symbol $\langle ... \rangle_{P(0,t)}$ stands for averaging with the
Wiener measure (\ref{wiener}) over all closed trajectories. The expression for
${\cal P}(S,L)$ can be obtained by making use of relation between this
quantity and density of states of a fictitious quantum particle with the mass
$m=\hbar(2D)^{-1}$ in a magnetic field: \cite{samokh}
\begin{equation}
{\cal P}(S,L)=\frac{\pi}{2l L}\cosh^{-2}\left(\frac{\pi S}{l L}\right).
\label{wsl}
\end{equation}
When obtaining Eq.\ (\ref{wsl}) we have used the expression for the diffusion
coefficient $D=v_F^2 \tau/2$, and normalized ${\cal P}(S,L)$ to unity:
\begin{equation}
\int_{-\infty}^{\infty} d S\, {\cal P}(S,L)=1. \label{normwsl}
\end{equation}
The function $w(S,L)$, which is connected with ${\cal P}(S,L)$ through the
diffusion density probability of return, by the following relationship
\begin{equation}
 w(S,L)=\frac{1}{4\pi D t}{\cal P}(S,L)=\frac{1}{2\pi l L}{\cal P}(S,L),
\label{WandP}
\end{equation}
is shown in Fig.\ \ref{fig41} for different $L$ values. In the same figure the
results of simulation are presented, too. As would be expected, the diffusion
theory describes only the statistics of long trajectories (see corresponding
curves labeled $50\, l$ and  $500\, l$ in Fig.\ \ref{fig41}). For ballistic
trajectories, there is no agreement between the simulation data and the above
theory.

To calculate the function $W(S)$, which is given by the Eq.\ (\ref{eq7a}), we
choose $l$ as the lower cutoff in the integral, because the ballistic
trajectories are not described within the diffusion approximation. Using Eqs.\
(\ref{wsl}) and (\ref{WandP}) we immediately obtain
\begin{equation}
4\pi l^2 W(S)=\frac{1}{S}\tanh\left(\frac{\pi S}{l^2} \right). \label{ws1}
\end{equation}
In Fig.\ \ref{fig3}, the results of calculations of the function $4\pi l^2
W(S)$ are presented by the dashed line. As is seen for large areas $S>l^2$,
the simple diffusion theory describes the simulation results perfectly. For
$S<l^2$ the theory gives the values of $W$ smaller than the simulation data.

In the framework of the diffusion approximation we have calculated the area
dependence of the average length of trajectories $\overline{L}$, introduced by
Eq.\ (\ref{eq7}). The results of numerical calculation are presented in Fig.\
\ref{fig4} by the dashed curve. It is evident that the theoretical and
simulation results are in good agreement only for $S>10\, l^2$.

Let us suggest some improvement of the above theory, which could allow one to
describe the statistic of ballistic trajectories as well. As is seen from
Fig.\ \ref{fig41}, at some value of $S=S_c$ (marked by arrows in Fig.\
\ref{fig41}) each curve obtained from the simulation procedure reveals a
step-like behaviour: there are no paths with $S>S_c$. Such a behaviour of
$w(S,L)$ is quite clear. A closed path with the length $L$ cannot enclose the
area larger than the area of a circle with the radius $L/(2\pi)$. This fact,
which has not been taken into account above, does not play an essential role
within the diffusion regime, because the value of $S_c$ is much greater than
the area enclosed by almost all the trajectories. In the ballistic regime,
$S_c$ is found to be equal to the areas enclosed by the most probable
trajectories (see Fig.\ \ref{fig41}, curves corresponding to $L=1\, l,\, 5\,
l$). Not counting the existence of $S_c$ in this case the function ${\cal
P}(S,L)$ (and $w(S,L)$ too) is found to be underestimated for $S<S_c$ owing to
the normalization of ${\cal P}(S,L)$ according to Eq.\ (\ref{normwsl}). The
artificial cutoff of the function ${\cal P}(S,L)$ obtained in the framework of
diffusion approximation (\ref{wsl}) (i.e. ${\cal P}(S,L)=0$ for
$|S|>L^2/(4\pi)$ ) and following normalization (\ref{normwsl}) of this
function gives
\begin{eqnarray}
 {\cal P}(S,L)&=&\frac{\pi}{2l
 L}\tanh^{-1}\left(\frac{L}{4l}\right)\times\nonumber \\
 &\times & \cosh^{-2}\left(\frac{\pi S}{l L}\right)
\Theta\left(\frac{L^2}{4\pi}-|S|\right)
 \label{wslImpr}
\end{eqnarray}
where $\Theta(x)$ is Heaviside step function. The use of this expression allows
to describe $w(S,L)$ behaviour for the ballistic trajectories much better than
Eq.\ (\ref{wsl}) (see the dash-dotted curve in Fig.\ \ref{fig41}). The results
of numerical calculation of $W(S)$ and $\overline{L}(S)$ obtained in the
framework of the improved theory are presented by the dash-dotted curves in
Figs.\ \ref{fig3} and \ref{fig4}, respectively. A good agreement with
simulation results is evident for $W(S)$ in the whole area range. However,
such an approach is too rough to describe $\overline{L}$-vs-$S$ curves .

Thus, the theoretical analysis shows that the simulation procedure works
correctly and allows to use it for analyzing, among other things, the
interference quantum corrections to the conductivity.

\section{Negative magnetoresistance} \label{sec:nmr}

Before the discussion of negative magnetoresistance let us consider the
results of calculation of $\delta\sigma$ for $b=0$. This will allow us to
realize the restrictions of our model in view of the finite size of matrix
used and limited number of collisions per one trajectory and to analyze for
what $\gamma$ range the simulation results can be applied to interpret
experimental data for macroscopic samples (when the sample dimensions are much
greater than the phase breaking length).

\begin{figure}
 \epsfclipon
 \epsfxsize=\linewidth
 \epsfbox{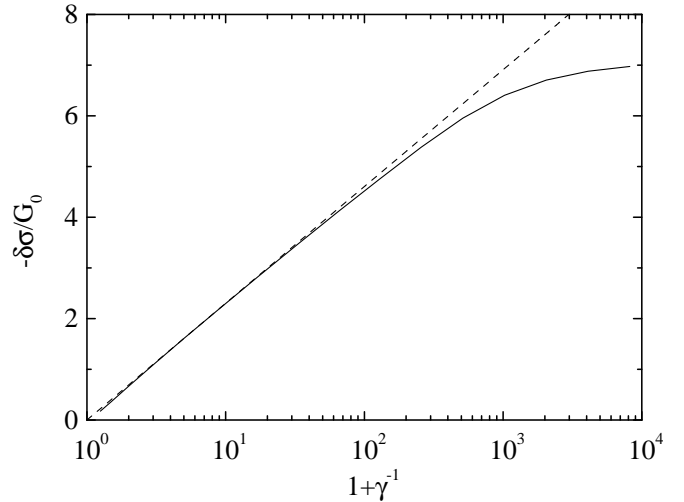}
\caption{The interference quantum correction to the conductivity in zero
magnetic field as a function of $\gamma$. The solid curve shows the simulation
results, the dashed line is the results of calculation according Eq.\
(\ref{qcbeq0}).
 } \label{fig42}
\end{figure}

To calculate $\delta\sigma$ we use Eq.\ (\ref{eq1}) assuming that each closed
path gives a contribution $1/I_s$ (where $I_s$ is a total number of paths) to
the probability of return. Each contribution is weighted by the factor $\exp
\left(- l_i/l_\varphi\right)$ to take into account the interference distortion
by inelastic processes. The final form of the expression for $\delta\sigma$ in
the case of $b=0$ looks as follows
\begin{equation}
\label{qcbeq0s} \frac{\delta\sigma}{G_0}=-\frac{2\pi l}{I_s d}\sum_{i} \exp
\left(- \frac{l_i}{l_\varphi} \right),
\end{equation}
where summation runs over all closed trajectories. Figure \ref{fig42} shows
the results of our simulation of $\delta\sigma$ as a function of
$\gamma=l/l_\varphi$ and, for comparison, the results of theoretical
calculation obtained through the well known exact formula
\begin{equation}\label{qcbeq0}
  \frac{\delta\sigma}{G_0}=\ln(1+\gamma^{-1}).
\end{equation}
As is seen for $\gamma\gtrsim 10^{-2}$ the simulation and theoretical data are
almost the same. A strong deviation is observed for $\gamma<10^{-3}$ where
just the trajectories with lengths $L\sim L_\varphi= \gamma^{-1}\, l> N_m\, l$
have to give an essential contribution to $\delta\sigma$ value. These
trajectories are not considered in our model. Thus, we believe that for
$\gamma=l/l_\varphi>10^{-2}$ our model system is equivalent to an unbounded 2D
system and the simulation gives correct results.

\subsection{Backscattering contribution}\label{subsec:back}
Now we are in position to discuss the magnetoconductance anomaly due to
suppression of quantum interference corrections by a magnetic field. To
calculate a magnetic field dependence of $\delta\sigma$ for our model system
we follow the ordinary way: the contribution of each closed trajectory to
$\delta\sigma$ is multiplied by the factor, which allows for the interference
distortion by the magnetic field. The final expression for $\delta\sigma(b)$
takes the form:
\begin{equation}
\label{eq4} \frac{\delta\sigma(b)}{G_0}=-\frac{2\pi l}{I_s d}\sum_{i}
\cos\left(\frac{(1+\gamma)^2 bS_i}{l^2}\right) \exp \left(-
\frac{l_i}{l_\varphi} \right).
\end{equation}
This formula is not meaningful from the experimental point of view, because
the absolute value of quantum correction is not a measurable quantity. The
magnetoconductance
$\Delta\sigma(b)=\sigma(b)-\sigma(0)=\delta\sigma(b)-\delta\sigma(0)$ is
usually  experimentally available.

\begin{figure}
 \epsfclipon
 \epsfxsize=\linewidth
 \epsfbox{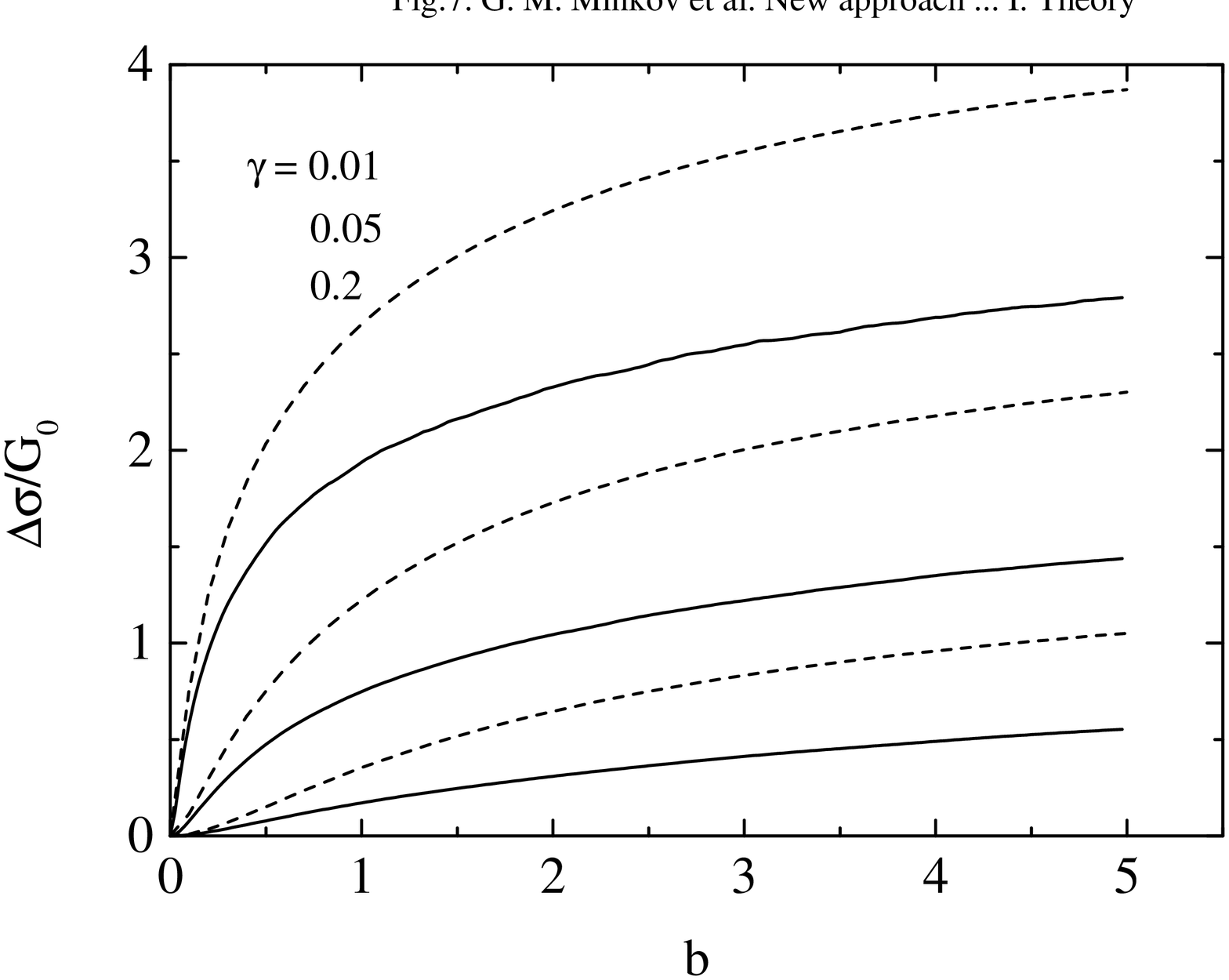}
  \caption{The magnetic field dependence of $\Delta\sigma$ for different
  $\gamma$ values. Solid curves are the results of simulation,  dashed
  curves are the diffusion limit (\ref{eq2}). Only backscattering contribution
  is taken into account.
  }\label{fig5}
\end{figure}

The results of calculation of $\Delta\sigma(b)$ for the model system are
presented in Fig.\ \ref{fig5} for several $\gamma$ values. They are in
excellent agreement with the results of numerical calculations carried out
beyond the diffusion approximation in Ref.\ \onlinecite{schm}. This lends
support to the correctness of parameters chosen for our model system: the set
of parameters used turns out to be suitable to simulate adequately
$\Delta\sigma(b)$ in the ranges of $b$ and $\gamma$ under consideration.
\begin{figure}
 \epsfclipon
 \epsfxsize=\linewidth
 \epsfbox{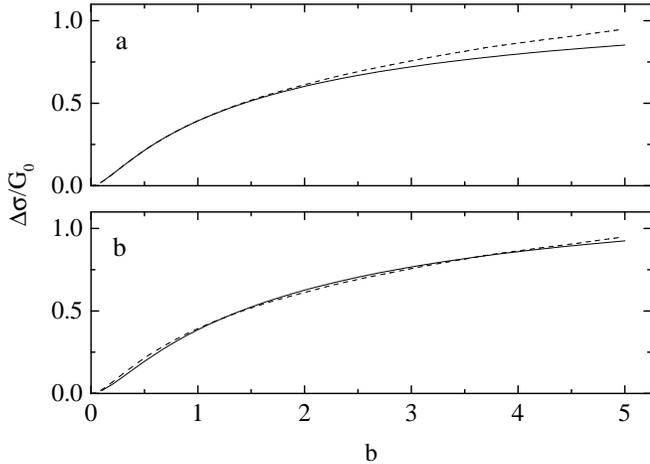}
\caption{The magnetic field dependence of $\Delta\sigma$ for $\gamma=0.1$.
Solid curves are the results of simulation. Dashed curves are the results of
fitting procedure with the use of Eq.\ (\ref{eq2}) made within two different
ranges of magnetic field: $b\leq 1$ (a) and $b\leq 5$ (b). The fitting
parameters are the following: $a=0.49$, $\gamma_f=0.087$ (a) and $a=0.65$,
$\gamma_f=0.12$ (b).}\label{fig6}
\end{figure}

In the same figure the results of calculation of $\Delta\sigma(b)$ in the
framework of diffusion approximation with the expression (\ref{eq2}) are
presented too. It is seen that this formula does not describe the simulated
$\Delta\sigma(b)$ curves. Even in the case of $\gamma=0.01$, when the
condition $\gamma\ll 1$ is seemingly fulfilled and the diffusion theory should
work well, the formula (\ref{eq2}) gives the value of $\Delta\sigma$
substantially larger than the simulation data: for $b=0.1$ the difference is
about $25 \%$.

As discussed above (see Section \ref{sec:itro}) Eq. (\ref{eq2}) is widely used
by experimenters to extract the values of $\gamma$ from experimental data. Two
fitting parameters are available in such a data processing: just as the value
of $\gamma$ so the prefactor $a$. To check the validity of this method in the
case when the diffusion approximation does not work, we have performed the
fitting procedure for $\Delta\sigma(b)$ curves simulated. In Fig.\ \ref{fig6}
the results of fitting procedure of $\Delta\sigma(b)$ made for $\gamma=0.1$
within two different ranges of magnetic field $b\leq 1$, and $b\leq 5$ are
presented by dotted and dashed curves, respectively. As is seen the simulated
curve is best described by Eq.\ (\ref{eq2}) in low magnetic field range.
However in both cases the ratio $\tau/\tau_\varphi$ found from the fitting
procedure and designated as $\gamma_f$ slightly differs from the value of
$\gamma$ used in the simulation. The difference is about $10 \%$ for $b\leq 1$
fitting range and $20\%$ for $b\leq 5$. The value of prefactor is found to be
significantly less than unity.

\begin{figure}
 \epsfclipon
 \epsfxsize=\linewidth
 \epsfbox{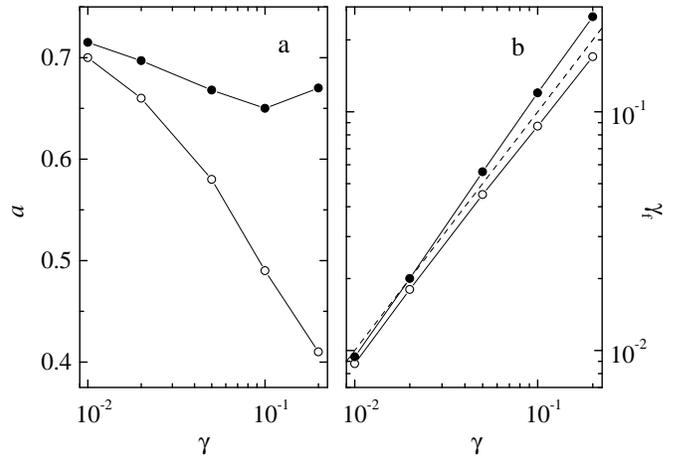}
\caption{The parameters $a$ (a) and $\gamma_f$ (b) as a function of $\gamma$
value fed into the simulation. Open and solid circles are the results of
fitting of $\Delta\sigma(b)$, obtained without non-backscattering
contribution, to Eq.\ (\ref{eq2}) in the ranges $b\leq 1$ and $b\leq 5$,
respectively. Solid lines are the guide for an eye, the dashed line shows
$\gamma=\gamma_f$ function.} \label{fig61}
\end{figure}

The results of such a data treatment for several $\gamma$ values are presented
in Fig.\ \ref{fig61}. It is clearly seen that the difference $10..30\%$
between $\gamma_f$ and $\gamma$ takes place for all $\gamma$ values. The value
of prefactor $a$ is always less than unity, and increases with decreasing
$\gamma$.

\subsection{Nonbackscattering contribution}\label{subsec:nonback}
\begin{figure}
 \epsfclipon
 \epsfxsize=\linewidth
 \epsfbox{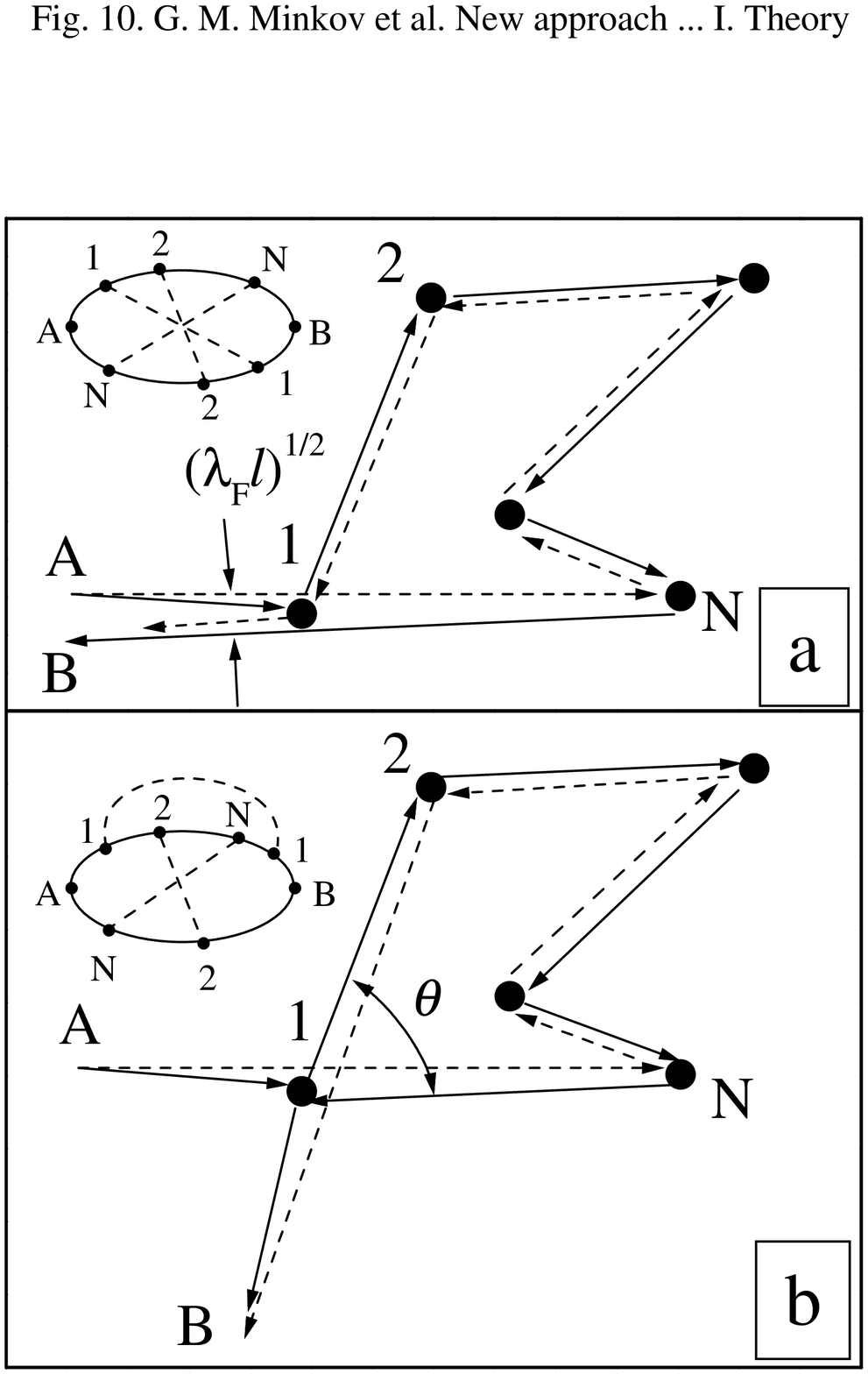}
\caption{Two types of coherent paths and corresponding diagrams relevant in
the first order in $(k_F l)^{-1}$, which are responsible for backscattering
(a) and nonbackscattering (b) contributions to weak localization, for given
configuration of scatterers.}\label{fig70}
\end{figure}

Up to this point we considered the coherent backscattering correction to
conductivity. The coherent paths for backscattering contribution are
schematically depicted in Fig.\ \ref{fig70}a. In this case the interference
takes place when the points $A$, $1$, $N$ and $B$ are close to one line.
However as it is shown in Ref.\ \onlinecite{dmit} there is one more variant of
coherent paths for the same configuration of scatterers, when the finish point
$B$ lies close to the line passing through the points $1$ and $2$ (see Fig.\
\ref{fig70}b). In the latter case the one of the two interfering waves is
scattered twice by the scatterer $1$ (which is regarded in this paper as a
starting point), while the other one passes the point $1$ without scattering.
This provides so called nonbackscattering contribution to conductivity. It was
shown in \cite{dmit} that both contributions are related to the probability of
return to the starting point for a given trajectory, the conductivity
correction due to nonbackscattering processes is positive. In our case such
processes can be easily taken into account by multiplying each term in Eq.\
(\ref{eq4}) by the factor $(1-\cos(\theta_i))$, where $\theta_i$ is the angle
between the first and last segments of the $i$-th closed path:
\begin{equation}\label{eq8}
\frac{\Delta\sigma(b)}{G_0}=-\frac{2\pi l}{I_s d}\sum_{i} \left[ ... \right]
\left( 1-\cos(\theta_i)\right).
\end{equation}

\begin{figure}
 \epsfclipon
 \epsfxsize=\linewidth
 \epsfbox{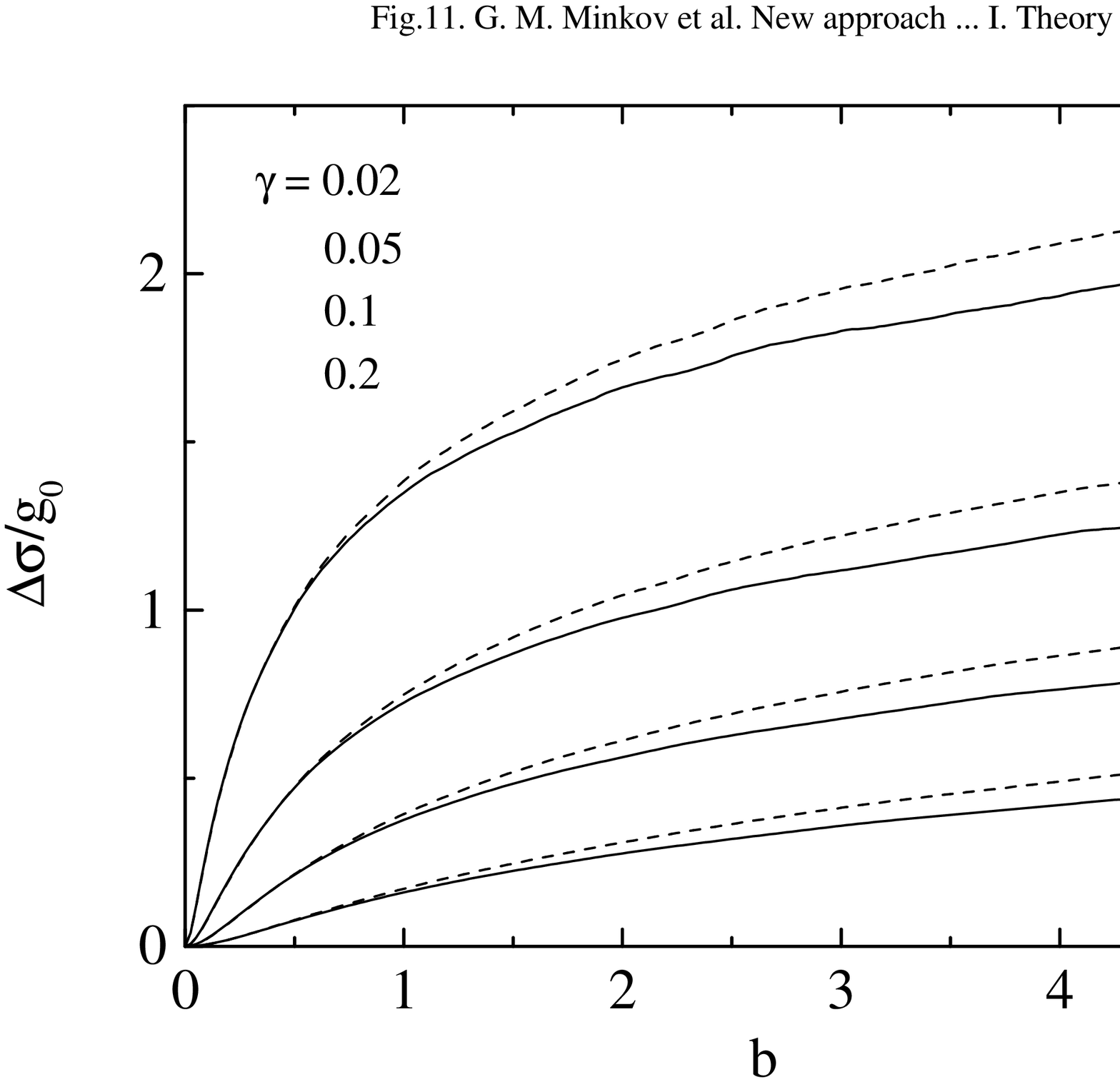}
\caption{The magnetic field dependence of $\Delta\sigma$ for different
$\gamma$ values. Dashed  and solid curves are the results of calculation
without and with non-backscattering contribution, respectively.}\label{fig7}
\end{figure}

Shown in Fig. \ref{fig7} are the magnetic field dependencies of
$\Delta\sigma$, calculated for different $\gamma$ values with the help of Eq.\
(\ref{eq8}). For comparison, the results computed with the formula (\ref{eq4})
are presented too. It is clearly seen that both types of calculations give
close results in the low magnetic field range $b\lesssim 0.5$. For higher
magnetic fields the inclusion of nonbackscattering contribution leads to a
decrease in magnetoconductance. Such a behaviour of the magnetoconductance is
in a good agreement with the results of numerical calculations presented in
Ref.\ \onlinecite{dmit}.

\begin{figure}
 \epsfclipon
 \epsfxsize=\linewidth
 \epsfbox{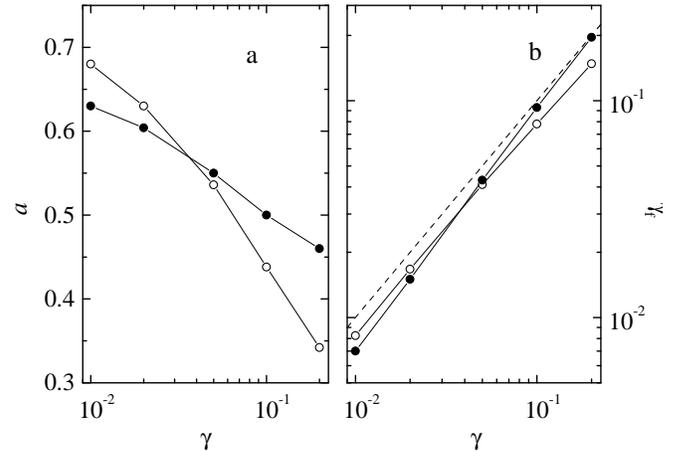}
\caption{The parameters $a$ (a) and $\gamma_f$ (b) as a function of $\gamma$
value fed into the simulation. Non-backscattering contribution is taken into
account. All designations are the same as in Fig.\ \ref{fig61}.} \label{fig71}
\end{figure}

As in the previous subsection we have attacked the simulated data as
experimental ones, i.e. $\Delta\sigma(b)$ curves have been fitted to the
calculated with Eq.\ (\ref{eq2}) values. The fitting results are presented in
Fig.\ \ref{fig71}. As one would expect, the values of $a$ and $\gamma_f$
obtained from the low magnetic field fitting ($b\leq 1$) are very close to
corresponding values in Fig.\ \ref{fig61}. This obviously results from the
fact that the solid and dashed curves in Fig.\ {\ref{fig7}} are close together
for these $b$ values. As for the parameters obtained from the whole magnetic
field range fitting, taking into account of the nonbackscattering contribution
results in decreasing both $a$ and $\gamma_f$ values.

%\bigskip
Thus, the standard method of experimental data processing with the use of Eq.\
(\ref{eq2}) allows to extract the phase breaking time (or length) with the
accuracy $(10-30)\,\%$ in wide ranges of $\gamma$ and $b$. The prefactor $a$
obtained this way is found to take the values from $0.3$ to $0.7$, i.e.
substantially less than unity. This reveals a possible reason of small value
of prefactor obtained from different experimental data treatment. This need
not be a consequence of an electron-electron interaction as discussed in some
papers, but can result from the fact that in real 2D systems the rigorous
conditions of diffusion approximation are not fulfilled even in the case of
rather small $\gamma$ and $b$.

\section{Conclusion}\label{sec:conc}
This paper is intended to demonstrate the possibility of obtaining the
information about the statistics of closed paths from the analysis of
anomalous magnetoconductance in 2D systems. In particular we have shown
explicitly that the statistic characteristics such as the area distribution
function of closed paths and area dependence of their average lengths
determine the magnetic field dependence of magnetoconductance. These functions
have been studied by using the computer simulation method. It has been shown
that in the ballistic regime the area distribution function of closed paths
deviates from $S^{-1}$ power law, which holds in the diffusion regime. The
theoretical analysis of simulation results has shown that such a behaviour of
$W(S)l$ is mainly connected with the existence of critical area value, which
corresponds to the maximal area enclosed by a path with a fixed length. It has
been shown that the area dependence of the average length of closed paths,
introduced through Eq.\ (\ref{eq7}), can be well described by the power
function, $\overline{L}\propto S^\beta$, with $\beta$ varying in the range
$0.55-0.62$, when $\gamma=\tau/\tau_\varphi$ changes from $0.1$ to $0.01$.

The results of simulation have been used to calculate the magnetoresistance of
the model 2D system. Both backscattering and nonbackscattering processes have
been taken into account. The calculated magnetic field dependencies of
$\Delta\sigma$ have been processed as experimental ones by a standard manner
with the help of Eq.\ (\ref{eq2}). It has been shown that the fitting
procedure gives $\tau/\tau_\varphi$ ratio, which differs from that used in the
simulation by the factor $0.8-1.3$. The value of prefactor obtained this way
is always less than unity.

\subsection*{Acknowledgments}
We thank A.P.~Dmitriev, V.Yu.~Kachorovskii, and A.G.Yashenkin for useful
discussions. This work was supported in part by the RFBR through Grants
97-02-16168, 98-02-17286, and 99-02-17110, the Russian Program {\it Physics of
Solid State Nanostructures} through Grant  97-1091, INTAS through Grant
97-1342 and the Program {\it University of Russia} through Grant 420.

\section*{Appendix}

For the sake of simplicity we suppose here that the start point coincides with
the origin. For the value of probability ${\cal T}$ that a trajectory passes
near the origin at the distance less than some prescribed value $d/2$ we can
write
\begin{equation}
{\cal T}=\sum_{N=3}^{N_m} {\cal T}_N, \label{eq1app}
\end{equation}
where ${\cal T}_{N}$ is the probability of return after $N$ collisions. The
value of ${\cal T}_{N}$ can be expressed through the probability density to
experience the ($N-1$)-th collision in the point $\bf{r}$, $W_{N-1}(\text{\bf
r})$:
\begin{equation}
{\cal T}_{N}= \int d \text{\bf r}\, {\cal P}(r)W_{N-1}(\text{\bf r}),
\label{eq2app}
\end{equation}
where ${\cal P}(r)$ is the probability that the particle moves without
collisions from the point
$\text{\bf r}$ to the circle of the radius $d/2$ around the
origin. It is easy to show that
\begin{equation}
{\cal P}(r)=\frac{1}{\pi}
\arctan\left(\frac{d}{2r}\right)\exp\left(-\frac{r}{l}\right). \label{eq3app}
\end{equation}
Then, for ${\cal T}_{N}$ we have
\begin{equation}
{\cal T}_{N}= \frac{1}{\pi}\int d\text{\bf r}\,
\arctan\left(\frac{d}{2r}\right)\exp\left(-\frac{r}{l}\right)
W_{N-1}(\text{\bf r}). \label{eq4app}
\end{equation}
For $d\ll l$, the inverse tangent in Eq.\ (\ref{eq4app})
can be replaced by its
argument. Analysis shows that for $d/l=10^{-2}$ (it is true for our case) this
gives an error in calculation of ${\cal T}_{N}$ less than $0.1$ \% for $N\ge
3$. Thus the expression for the probability ${\cal T}$ becomes
\begin{equation}
{\cal T}\simeq\frac{d}{2\pi}\sum_{N=3}^{N_m} \int \frac{d\text{\bf
r}}{r}\exp\left(-\frac{r}{l}\right) W_{N-1}(\text{\bf r}). \label{eq5app}
\end{equation}
Comparing Eq.\ (\ref{eq5app}) with the expression for the probability density
$W$ \cite{dyak,schm}
\begin{equation}
W=\frac{1}{2\pi l}\sum_{N=3}^{N_m} \int \frac{d\text{\bf
r}}{r}\exp\left(-\frac{r}{l}\right) W_{N-1}(\text{\bf r}), \label{eq6app}
\end{equation}
we can conclude that
\begin{equation}
W\simeq (ld)^{-1}{\cal T}. \label{eq7app}
\end{equation}

\end{multicols}

\begin{references}
\bibitem{otsenka}The bending of the trajectory in a magnetic field due to
Lorentz force is neglected provided that the cyclotron radius is much greater
than the mean free path. This condition corresponds to the requirement $k_F
l/b\gg 1$. Since all the theory of weak localization is true when $k_F l\gg 1$
the above condition is well justified in the whole range of magnetic fields
under study throughout this paper ($b<5$).
\end{references}
\end{document}